\documentclass[12pt]{article}\usepackage[]{graphicx}\usepackage[]{color}
\usepackage[margin=1in]{geometry}
\usepackage{amsmath}
\usepackage{bm}
\usepackage[numbers]{natbib}
\usepackage[compact]{titlesec}

\usepackage[english]{babel}
\usepackage{multirow}
\usepackage{indentfirst}
\usepackage[doublespacing]{setspace}
\usepackage{lineno}
\usepackage{todonotes}
\usepackage[hidelinks]{hyperref}
\usepackage{array}
\usepackage{ragged2e}
\usepackage{float}
\usepackage{caption}

\titleformat*{\section}{\large\bfseries}
\titleformat*{\subsection}{\bfseries}
\IfFileExists{upquote.sty}{\usepackage{upquote}}{}

\begin{document}
\begin{center}

{\huge
Evaluating the Impact of Instrumental Variables in Propensity Score Models Using Synthetic and Negative Control Experiments\\[5mm]
}

{\Large
Yuxi Tian\textsuperscript{1}, 
Nicole Pratt\textsuperscript{2}, 
Laura L.~Hester\textsuperscript{3}, \\
George Hripcsak\textsuperscript{4,5},
Martijn J.~Schuemie\textsuperscript{3,6}, 
Marc A.~Suchard\textsuperscript{1,6,7} \\[5mm] 
}
\end{center}

\begin{enumerate}
\item Department of Computational Medicine, David Geffen School of Medicine at UCLA, University of California, Los Angeles, CA, USA 90095
\item Quality Use of Medicines and Pharmacy Research Centre, Sansom Institute for Health Research, University of South Australia GPO Box 2471, Adelaide, South Australia 5001, Australia
\item Epidemiology Analytics, Janssen Research and Development LLC, Titusville, NJ, USA 08560
\item Department of Biomedical Informatics, Columbia University Medical Center, New York, NY, USA 10032
\item Medical Informatics Services, NewYork-Presbyterian Hospital, New York, NY, USA, 10032
\item Department of Biostatistics, UCLA Fielding School of Public Health, University of California, Los Angeles, CA, USA 90095
\item Department of Human Genetics, David Geffen School of Medicine at UCLA, University of California, Los Angeles, CA, USA 90095
\end{enumerate}

\clearpage

\section*{Abstract}
In pharmacoepidemiology research, instrumental variables (IVs) are variables that strongly predict treatment but have no causal effect on the outcome of interest except through the treatment.
There remain concerns about the inclusion of IVs in propensity score (PS) models amplifying estimation bias and reducing precision.
Some PS modeling approaches attempt to address the potential effects of IVs, including selecting only covariates for the PS model that are strongly associated to the outcome of interest, thus screening out IVs.
We conduct a study utilizing simulations and negative control experiments to evaluate the effect of IVs on PS model performance and to uncover best PS practices for real-world studies.
We find that simulated IVs have a weak effect on bias and precision in both simulations and negative control experiments based on real-world data.
In simulation experiments, PS methods that utilize outcome data, including the high-dimensional propensity score, produce the least estimation bias.
However, in real-world settings underlying causal structures are unknown, and negative control experiments can illustrate a PS model's ability to minimize systematic bias.
We find that large-scale, regularized regression based PS models in this case provide the most centered negative control distributions, suggesting superior performance in real-world scenarios.

\section{Introduction}
Propensity scores (PS), an estimate of treatment assignment probability, are widely used for confounding control in observational studies \cite{rubin1997estimating, rosenbaum1983central}.
PS adjustment allows for the comparison of similar treated and control persons in a cohort, through methods that group individuals with similar PS together for analysis \cite{rubin2007design}.
There remains controversy over the issue of variable selection for the PS model in high-dimensional datasets where the number of covariates can range from the hundreds to tens of thousands.
Traditionally, clinical investigators construct a PS model using expert domain knowledge, including only covariates known or suspected to the investigators as confounders.
However, this human-dependent process can be substantially and inexorably biased \cite{king2015propensity}.

Various automated procedures for PS model selection exist to eliminate human bias from the task of selecting a parsimonious PS model out of thousands of available covariates.
Still, concern remains over whether to include all pretreatment covariates in the automated selection process or whether to first curate them to only include ``real" confounders that will ultimately reduce estimation bias in the outcome of interest.
In particular, there are concerns over instrumental variables (IVs) that causally affect the outcomes only through their effect on the treatment \cite{greenland2000introduction}.
Instrumental variables are covariates that are associated with treatment, independent of all confounders, and independent of the outcome conditional on treatment and confounders \cite{greenland2000introduction}.
When present, IVs themselves can actually be used to estimate average treatment effects, and provide unbiased bounds on the treatment effect size \cite{angrist1991instrumental,angrist1996identification,martens2006instrumental}.
``IVs" also more broadly refer to variables that meet the mentioned criteria that can be included in a PS model.

When used as conditioning variables in a PS model, IVs  can be the source of so called ``Z-bias," whereby they may increase bias from unmeasured confounders in observational data \cite{brookhart2010confounding,brooks2013squeezing}.
As such, IVs are also known as ``bias amplifiers" for amplifying existing residual bias after conditioning on other measured confounders \cite{pearl2011invited,ding2017instrumental}.
The potential deleterious effects of IVs have been shown in both theoretical frameworks \cite{ding2017instrumental,wooldridge2016should} and small simulation studies \cite{bhattacharya2007instrumental,lefebvre2008impact,myers2011effects}.
In addition to amplifying bias, conditioning on IVs may also reduce precision \cite{brookhart2006variable,caruana2015new}.

While IVs can be easily simulated, it is controversial how prevalent IVs are and how to identify them in real-world data.
IVs are sensitive to deviations from their unverifiable definitional assumptions \cite{martens2006instrumental,hernan2006instruments}, and perfect IVs are difficult to identify for IV analyses \cite{d2007estimating}.
In real-world observational health data, researchers often use provider characteristics as IVs, such as distance to health care facility or physician variation \cite{garabedian2014potential}, but these IVs can be flawed and also unavailable in large-scale insurance claims databases that are used for many observational studies.
In the absence of tools for identifying quality IVs, there is a movement to only include covariates associated with the outcome in propensity score models \cite{patrick2011implications}.
The popular high-dimensional propensity score (HDPS) \cite{schneeweiss2009high} selects only covariates that have a high apparent relative risk with the outcome \cite{bross1966spurious}.

In this study, we conduct simulations and negative control experiments to explore the effect of IVs in real-world data and optimal PS models for reducing bias in the presence of IVs.
We base our simulations on real-world data, resulting in much larger models than those used in existing simulation studies.
We also explore calendar year as a potential IV that would be readily available in longitudinal data, an approach that has prior practice \cite{cain2009effect,mack2015comparative}.
Our compared PS models contrast the approach of selecting model covariates for treatment prediction versus outcome association.
In addition to reporting the simulation bias and precision, we introduce the use of negative control outcomes to measure the effects of including an IV in the PS model on residual bias \cite{lipsitch2010negative,arnold2016negative}.

\section{Methods}

\subsection{Clinical study}

We base our experiments on a real-world anticoagulants study of first time dabigatran to warfarin users among patients with atrial fibrillation from 2010 - 2018 using the IBM MarketScan Medicare Supplemental and Coordination of Benefits Database.
The primary outcome is gastrointestinal bleeding.
This is a reproduction of a published observational study \cite{graham2014cardiovascular} and follows a new user cohort study design \cite{ray2003evaluating,ryan2013empirical}.
We use the \textsc{CohortMethod} R package \cite{schuemie2017cohortmethod} to construct the study cohort.

\subsection{PS models}

In addition to an unadjusted analysis, we experiment with several PS models, all of which are fit using large-scale, $L_1$ regularized regression models \cite{tian2018evaluating} through the \textsc{Cyclops} R package \cite{suchard2013massive}.
We use 10-fold cross-validation to allow the data to choose the strength of regularization.
Five of these PS models do not included simulated IVs.
Firstly, we use all measured covariates (All Covariates) in the PS model to maximize treatment prediction; these covariates age, sex, calendar year, all prior diagnoses, drugs, and procedures.
Secondly, we examine the fitted All Covariates PS model and select the calendar year covariate with the largest absolute coefficient, which is the indicator for 2010.
We then exclude 2010 from the set of covariates for PS estimation.
This indicator for 2010 is a strong predictor for warfarin because dabigatran was just coming onto the market at that time, and physician awareness and preference could have been a factor in dabigatran use.
We fit a large-scale, $L_1$ regularized Cox proportional hazards outcome model for GI bleed, and find that calendar year has a zero coefficient after cross-validation, satisfying the IV assumption of no association with the outcome.
Thirdly, we exclude \textit{all} calendar year indicators from the All Covariates model.
Fourthly, we screen the most prevalent 500 covariates according to the HDPS apparent-relative-risk criterion and only include in the PS model the 200 top ranked covariates (according to apparent relative risk).
We call this set of 200 covariates the ``HDPS Set."
Fifthly, we use the fitted Cox outcome model for gastrointestinal and include in the PS model only the covariates that have non-zero coefficients.
These covariates have the highest conditional association with the outcome in a multivariate model, and we call this covariate set the ``Cox Set."

\begin{figure}[H]
	\centering
  	\captionsetup{width=\textwidth}
  	\includegraphics[width=\textwidth] {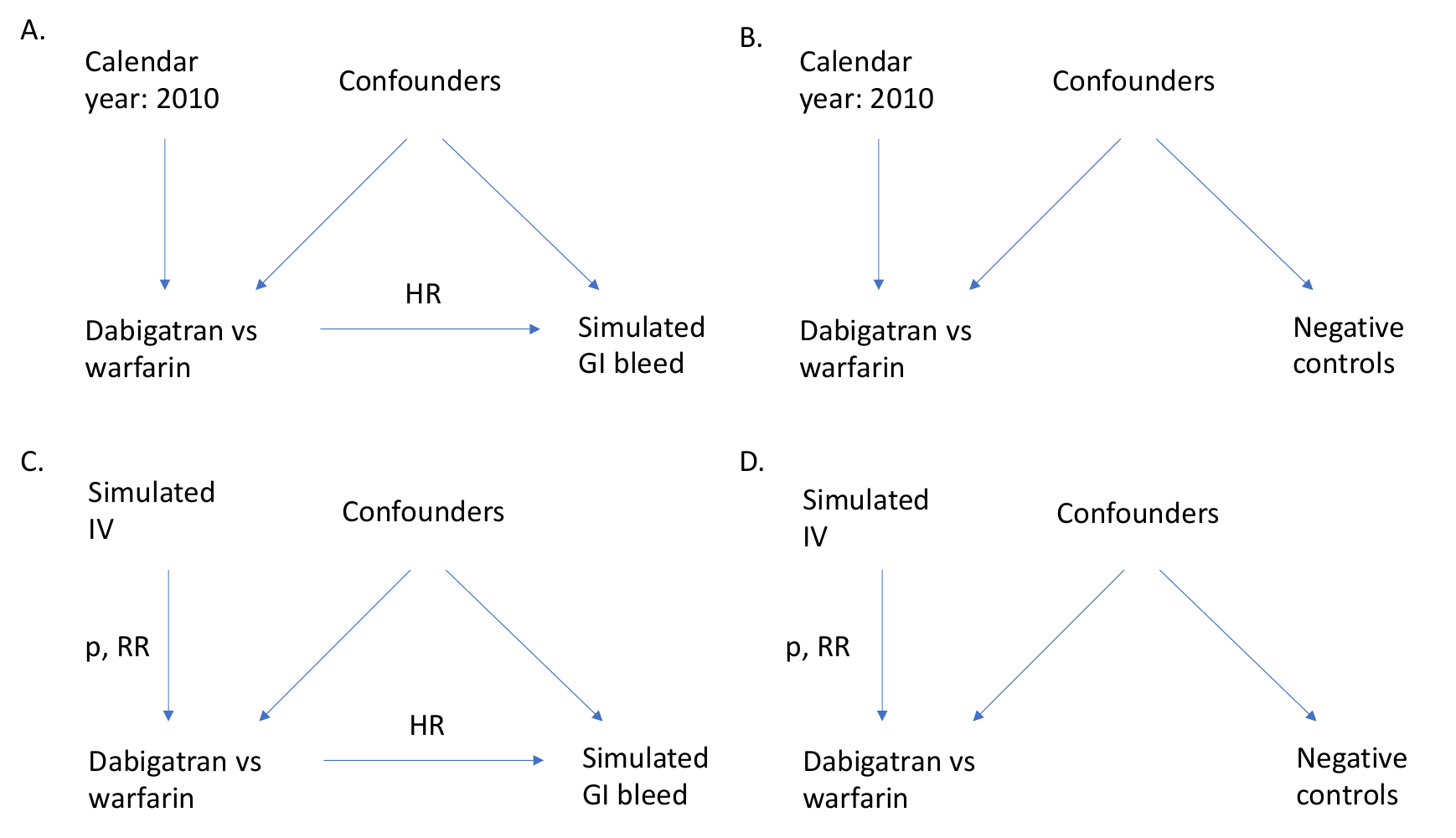}
  	\caption{A) directed acyclic graph (DAG) showing calendar year 2010 as an IV affecting estimation of hazard ratio (HR) of simulated outcomes of GI bleed. B) DAG showing negative control outcomes with no presumed effect from compared treatments. C) DAG showing effects of simulated IV under specified prevalence and relative ratio on the HR estimate of simulated outcome. D) DAG showing effects of simulated IV on negative control estimates}
	\label{fig: IVsimulations}
\end{figure}

The above PS models (models 1 - 6 in Table \ref{table: ps_models}) are fit only once, whereas the PS models which include a simulated IV (models 7-33 in Table \ref{table: ps_models}) and are fit once per simulation.
We use three baseline covariate sets corresponding to models 2, 5, 6 in Table \ref{table: ps_models}.
To these PS models, we add a single simulated IV with one of three prevalences ($p = 0.025\%, 0.05\%, 0.1\%$), and one of three relative risks with the treatment variable ($RR = 1.5, 2, 4$).
For each of the three baseline covariate sets, there are 9 additional PS models, one for each prevalence-relative risk combination.
There are a net total of 33 PS models, listed in Table \ref{table: ps_models}.
DAGs representing simulations using simulated outcomes are shown in Figures \ref{fig: IVsimulations}A and  \ref{fig: IVsimulations}C.

\begin{table}[H]
\centering
\begin{tabular}{|l|l|l|l|}
  \hline
  1. Unadjusted 		& 7. All + 0.025/1.5 	& 16. HDPS + 0.025/1.5 	& 25. Cox + 0.025/1.5  \\
  2. All Covariates 	& 8. All + 0.025/2 	& 17. HDPS + 0.025/2 	& 26. Cox + 0.025/2 \\
  3. No 2010 		& 9. All + 0.025/4 	& 18. HDPS + 0.025/4 	& 27. Cox + 0.025/4 \\
  4. No Years 		& 10. All + 0.05/1.5 	& 19. HDPS + 0.05/1.5 	& 28. Cox + 0.05/1.5 \\
  5. HDPS Set 		& 11. All + 0.05/2 	& 20. HDPS + 0.05/2 	& 29. Cox + 0.05/2 \\
  6. Cox Set 		& 12. All + 0.05/4 	& 21. HDPS + 0.05/4 	& 30. Cox + 0.05/4 \\
  				& 13. All + 0.1/1.5 	& 22. HDPS + 0.1/1.5 	& 31. Cox + 0.1/1.5 \\
  				& 14. All + 0.1/2 	& 23. HDPS + 0.1/2 		& 32. Cox + 0.1/2 \\
  				& 15. All + 0.1/4 	& 24. HDPS + 0.1/4 		& 33. Cox + 0.1/4 \\
  \hline
   \end{tabular}
\caption{Evaluated PS models. Simulated IVs have prevalence $p$ and relative risk with treatment $r$ represented as $p/r$. Models 7-15 add a simulated IV to the Model 2. Models 16-24 add a simulated IV to Model 5. Models 25-33 add a simulated IV to Model 6.}
\label{table: ps_models}
\end{table}

\subsection{Outcome simulations}

Using the real-world anticoagulants study as a simulation framework, we simulate new outcomes according to a ``plasmode" design \cite{franklin2014plasmode,vaughan2009use}.
The specific plasmode design we employ is detailed in \cite{tian2018evaluating}.
We first fit the full cohort data to a Cox proportional hazards model to obtain a realistic survival model for outcome simulation, including covariate coefficients and survival functions for the outcome and censoring events.
Keeping the original covariates and covariate data, we calculate each subject's linear predictor, and simulate outcome times and censoring times under a desired true hazard ratio.
We simulate under four true hazard ratios (1, 1.5, 2, 4).
With each fitted PS model, we perform variable length PS matching \cite{austin2008assessing}  with a maximum ratio of 10:1 and a caliper of 0.2, and use variable-ratio matching \cite{rassen2012one}.
We then fit a PS-stratified Cox proportional hazard model with the simulated outcomes, to obtain point estimates and 95\% confidence intervals of the treatment effect size.

\subsection{Negative controls}
Negative control outcomes are an emerging tool in observational research that allow for experiments using real data by providing a standard of clinical truth -- that of no effect between exposure and outcome \cite{lipsitch2010negative,arnold2016negative}.
In a observational clinical study setting, a negative control is an outcome that investigators can determine, with some confidence, is not differentially affected by the active treatment or reference treatment.
Effect estimation on a large set of negative control outcomes provides a distribution whose deviation from the expected null effect approximates the systemic study bias, or residual bias after controlling for measured confounding \cite{schuemie2014interpreting}.

We identify 49 negative control outcomes for dabigatran versus warfarin through a data-rich algorithm \cite{voss2017accuracy} combined with manual curation.
Further details on negative controls were selected are given in the Supplementary Material.
Similar to the simulated outcomes, we perform variable ratio PS matching and fit PS-stratified Cox proportional hazards models for the negative control outcomes.
We fit the set of negative control estimates -- each presumed to have a true hazard ratio of 1 -- to an empirical null distribution \cite{schuemie2014interpreting}.
This distribution characterizes the study residual bias after PS adjustment \cite{schuemie2018empirical} and arises from both unmeasured confounding and inappropriate control of measured confounding, such as inclusion of IVs.
DAGs representing simulations using negative control outcomes are shown in Figures \ref{fig: IVsimulations}B and  \ref{fig: IVsimulations}D.

\subsection{Metrics}
For both simulated and negative control outcomes, we compare the bias and standard deviation (SD) of the estimated hazard ratios for dabigatran versus warfarin to the true hazard ratios (known for the simulated outcomes and presumed to be 1 for the negative control outcomes).
We fit the negative control estimates to empirical null distributions, and report the distribution means and SDs.
To assess how the IVs affect the covariate balance through the PS, we compare the standardized mean differences (SMDs) for all covariates before and after PS matching.
We plot the distribution of SMDs and also note the number of after-matching SMDs that cross a threshold of 0.1 \cite{austin2009balance}.
We also plot the distributions of the fitted PS models.

\section{Results}

In the anticoagulants study, there are 20,474 first-time dabigatran users and 56,648 first-time warfarin users.
There are 52,729 total unique covariates in the All Covariates set, of which 900 have nonzero coefficients in the fitted PS model.
The HDPS Set of covariates contains the 200 covariates with the highest apparent relative risk out of the 500 covariates with the highest prevalence.
Of these, 170 have nonzero coefficients in the fitted PS model.
The Cox Set of covariates obtained from a large-scale outcome model contains 74 covariates, and 73 of them have nonzero coefficients in the fitted PS model.
There are 31 covariates that overlap between the PS model for All Covariates and the Cox PS model, 26 that overlap between the HDPS PS model and the Cox PS model, and 93 that overlap between the All Covariates PS model and the HDPS PS model.

\begin{figure}[H]
	\centering
  	\captionsetup{width=\textwidth}
  	\includegraphics[width=\textwidth, height=1.25\textwidth] {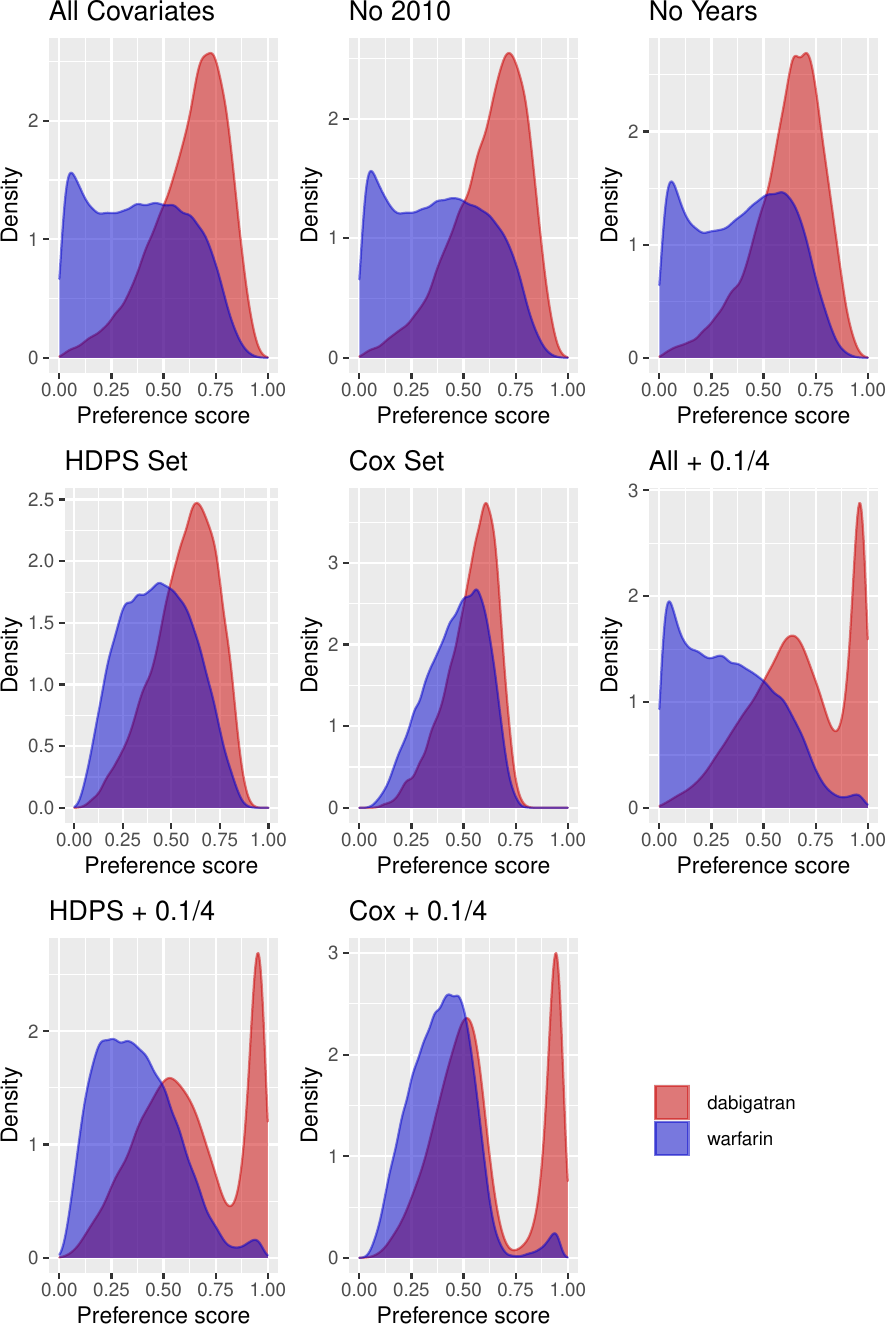}
  	\caption{Propensity score distributions represented as preference scores that normalize propensity scores by prevalence. The three plots from simulated IV PS models are taken from simulations with 10\% IV prevalence and relative risk of 4.}
	\label{fig: ps_plots}
\end{figure}

Figure \ref{fig: ps_plots} plots for the PS models the preference score distributions that normalize propensity scores by the treatment prevalences \cite{walker2013tool}.
For all PS methods, at least 68\% of subjects lie in a region of equipoise between 0.3 and 0.7 preference score.
There are few discernible differences among the PS plots for All Covariates and All Covariates with 2010 removed and with all calendar years removed.
These three PS distributions show moderately strong differentiation between the dabigatran and warfarin populations.
In contrast, the PS distributions built on the HDPS Set and Cox Set of covariates show less differentiation between distributions.
All three distributions with a simulated IV have large spikes in the dabigatran distribution close to 1, showing that the simulated IV has a strong discriminatory effect on the PS distributions.
On inspection of the PS models over 100 simulations, every single simulated PS model includes the simulated IV as a covariate with nonzero coefficient.

\begin{figure}[H]
	\centering
  	\captionsetup{width=\textwidth}
  	\includegraphics[width=\textwidth, height=1.25\textwidth] {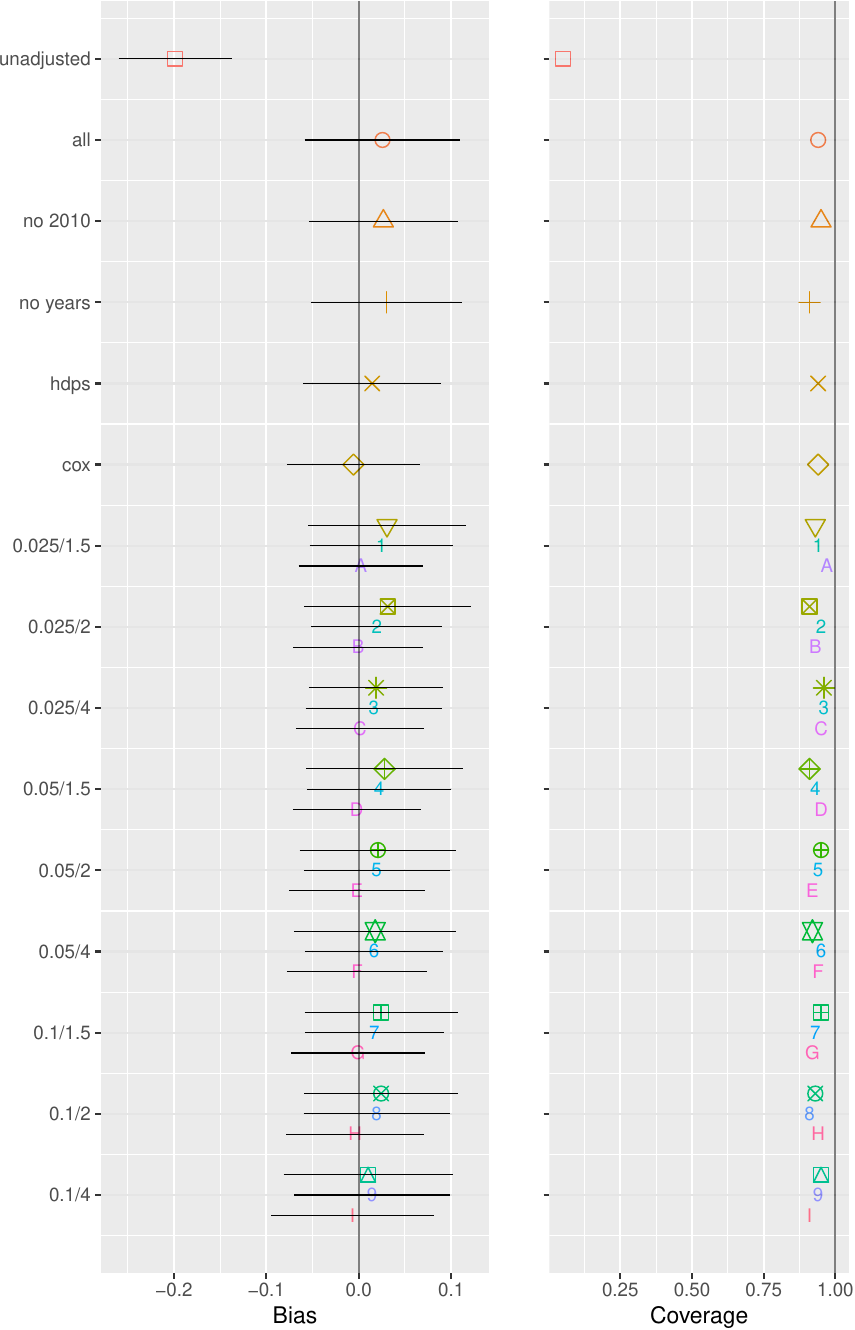}
  	\caption{Left: bias and SD of simulation experiments with true hazard ratio of 4. Right: coverage of true effect size of $HR=4$ across 100 simulations. For the 9 simulated IV settings, the shapes represent All Covariates, the numbers represent HDPS set, and the letters represent Cox set.}
	\label{fig: sim_bias4}
\end{figure}

Simulation results under a true hazard ratio of 4 are shown in Figure \ref{fig: sim_bias4}, with mean and SD provided in Table \ref{table: bias4}.
The unadjusted estimate has by far the largest bias and lowest coverage of the true effect size of all compared methods.
Relative to All Covariates, removing calendar year 2010 and all calendar years only very slightly increases the bias, and has minimal effect on the variance.
The HDPS Set has smaller bias and variance compared to All Covariates, and the Cox Set has almost no bias and even smaller variance.
For simulations with simulated IV based on All Covariates and HDPS Set, increasing the simulated IV prevalence and relative risk actually decreases the study bias, while having limited effect on the variance.
The simulations with simulated IV added to Cox Set has the smallest bias, followed by those added to HDPS Set, then those added to All Covariates.
All methods other than unadjusted have high coverage of the true effect size across the 100 simulations.
Simulation results under the other three simulated true hazard ratios display similar patterns and are shown in the Supplementary Material.

\begin{table}[H]
\centering
\begin{tabular}{|l|l|l|l|}
  \hline
  1. Unadjusted 		&  -0.199 (0.061) 	& 16. HDPS + 0.025/1.5 	&  0.024 (0.078) \\
  2. All Covariates 	&  0.026 (0.084) 	& 17. HDPS + 0.025/2 	&  0.020 (0.071) \\
  3. No 2010 		&  0.027 (0.081) 	& 18. HDPS + 0.025/4 	&  0.016 (0.074) \\
  4. No Years 		&  0.030 (0.082) 	& 19. HDPS + 0.05/1.5 	&  0.022 (0.078) \\
  5. HDPS Set 		&  0.015 (0.075) 	& 20. HDPS + 0.05/2 	&  0.019 (0.079) \\
  6. Cox Set 		&  -0.006 (0.072)	& 21. HDPS + 0.05/4 	&  0.017 (0.075) \\
 				&  				& 22. HDPS + 0.1/1.5 	&  0.017 (0.075) \\
 				& 				& 23. HDPS + 0.1/2 		&  0.019 (0.079) \\
				& 				& 24. HDPS + 0.1/4 		&  0.014 (0.084) \\
 7. All + 0.025/1.5	& 0.031 (0.086)	    	& 25. Cox + 0.025/1.5	& 0.002 (0.067) \\
 8. All + 0.025/2		& 0.031 (0.091) 	& 26. Cox + 0.025/2		& -0.001 (0.070) \\
 9. All + 0.025/4		& 0.019 (0.073)	    	& 27. Cox + 0.025/4		& 0.001 (0.069) \\
 10. All + 0.05/1.5	& 0.028 (0.085)	    	& 28. Cox + 0.05/1.5		& -0.002 (0.069) \\
 11. All + 0.05/2		& 0.021 (0.085)	    	& 29. Cox + 0.05/2		& -0.002 (0.073) \\
 12. All + 0.05/4		& 0.018 (0.088)	    	& 30. Cox + 0.05/4		& 0.002 (0.076) \\
 13. All + 0.1/1.5	& 0.024 (0.083)	    	& 31. Cox + 0.1/1.5		& -0.001 (0.072) \\
 14. All + 0.1/2		& 0.024 (0.083)	    	& 32. Cox + 0.1/2		& -0.004 (0.074) \\
 15. All + 0.1/4		& 0.010 (0.091)	    	& 33. Cox + 0.1/4		& -0.007 (0.088) \\
  \hline
   \end{tabular}
\caption{Simulation bias for true $HR = 4$, as Mean (SD), for all PS models}
\label{table: bias4}
\end{table}

\begin{figure}[H]
	\centering
  	\captionsetup{width=\textwidth}
  	\includegraphics[width=\textwidth, height=1.25\textwidth] {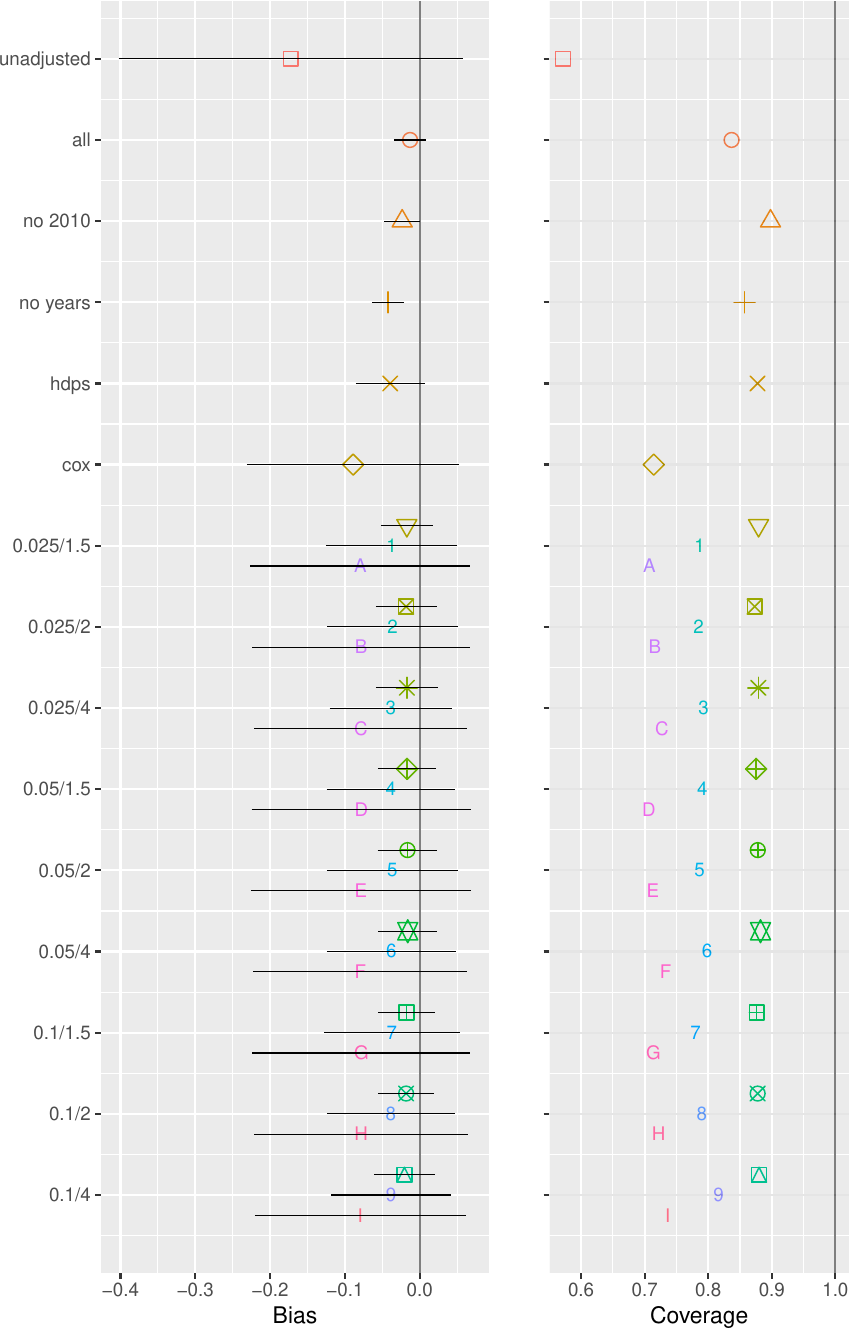}
  	\caption{Left: mean and SD of fitted negative control distributions, characterizing residual study bias. Right: coverage of presumed true effect size of 1 HR by negative control estimates. For the 9 simulated IV settings, the shapes represent All Covariates, the numbers represent HDPS set, and the letters represent Cox set.}
	\label{fig: neg_bias}
\end{figure}

While the above results represent performance under a known outcome model with only measured confounding, the negative control distributions approximate the residual study bias in real-world settings.
The null distribution means and SD are shown in Figure \ref{fig: neg_bias} and Table \ref{table: neg_dist}.
The unadjusted estimate has by far the largest bias (deviation from 0 mean) and variance, and the lowest coverage of unity HR by the individual negative control estimates.
Among PS models without simulated IVs, All Covariates has the smallest bias, while removing calendar year 2010 and all calendar years creates increasingly larger bias but also greater precision.
HDPS Set has larger bias and variance than All Covariates, though higher coverage.
Meanwhile, Cox Set has even larger bias and variance than HDPS Set and lower coverage than all other PS models.
Among PS models with a simulated IV, for each IV prevalence and relative risk the All Covariates estimate has smaller bias and variance and lower coverage than the HDPS Set estimate, which in turn has smaller bias and variance and lower coverage than the Cox Set estimate.
Adding an instrumental variable to All Covariates marginally increases the bias and variance, and also the coverage, though the magnitude of increase is not clearly associated with the strength of the IV.
However, adding an IV to the HDPS Set slightly decreases the bias overall, and noticeably increases the variance and lowers the coverage.
Finally, adding an IV to the Cox Set decreases the bias and variance, and manages to increase the coverage under some settings.
Increasing the relative risk of the simulated IV slightly increases the coverage throughout.

\begin{table}[H]
\centering
\begin{tabular}{|l|l|l|l|}
  \hline
  1. Unadjusted 		&  -0.173 (0.229) 	& 16. HDPS + 0.025/1.5 	&  -0.038 (0.088) \\
  2. All Covariates 	&  -0.013 (0.028) 	& 17. HDPS + 0.025/2 	&  -0.036 (0.087) \\
  3. No 2010 		&  -0.023 (0.024) 	& 18. HDPS + 0.025/4 	&  -0.039 (0.082) \\
  4. No Years 		&  -0.042 (0.017) 	& 19. HDPS + 0.05/1.5 	&  -0.039 (0.086) \\
  5. HDPS Set 		&  -0.039 (0.051) 	& 20. HDPS + 0.05/2 	&  -0.037 (0.088) \\
  6. Cox Set 		&  -0.089 (0.147)	& 21. HDPS + 0.05/4 	&  -0.038 (0.086) \\
 				&  				& 22. HDPS + 0.1/1.5 	&  -0.037 (0.091) \\
 				& 				& 23. HDPS + 0.1/2 		&  -0.038 (0.085) \\
				& 				& 24. HDPS + 0.1/4 		&  -0.038 (0.080) \\
 7. All + 0.025/1.5	& -0.017 (0.035)	& 25. Cox + 0.025/1.5	& -0.079 (0.147) \\
 8. All + 0.025/2		& -0.018 (0.041) 	& 26. Cox + 0.025/2		& -0.078 (0.146) \\
 9. All + 0.025/4		& -0.017 (0.041)	& 27. Cox + 0.025/4		& -0.079 (0.142) \\
 10. All + 0.05/1.5	& -0.017 (0.039)	& 28. Cox + 0.05/1.5		& -0.078 (0.147) \\
 11. All + 0.05/2		& -0.016 (0.040)	& 29. Cox + 0.05/2		& -0.079 (0.147) \\
 12. All + 0.05/4		& -0.016 (0.039)	& 30. Cox + 0.05/4		& -0.080 (0.143) \\
 13. All + 0.1/1.5	& -0.018 (0.038) 	& 31. Cox + 0.1/1.5		& -0.078 (0.145) \\
 14. All + 0.1/2		& -0.018 (0.038)	& 32. Cox + 0.1/2		& -0.078 (0.143) \\
 15. All + 0.1/4		& -0.020 (0.040)	& 33. Cox + 0.1/4		& -0.079 (0.141) \\
  \hline
   \end{tabular}
\caption{Mean and (SD) of fitted negative control distributions for all PS models}
\label{table: neg_dist}
\end{table}

\begin{figure}[H]
	\centering
  	\captionsetup{width=\textwidth}
  	\includegraphics[width=\textwidth, height=1.25\textwidth] {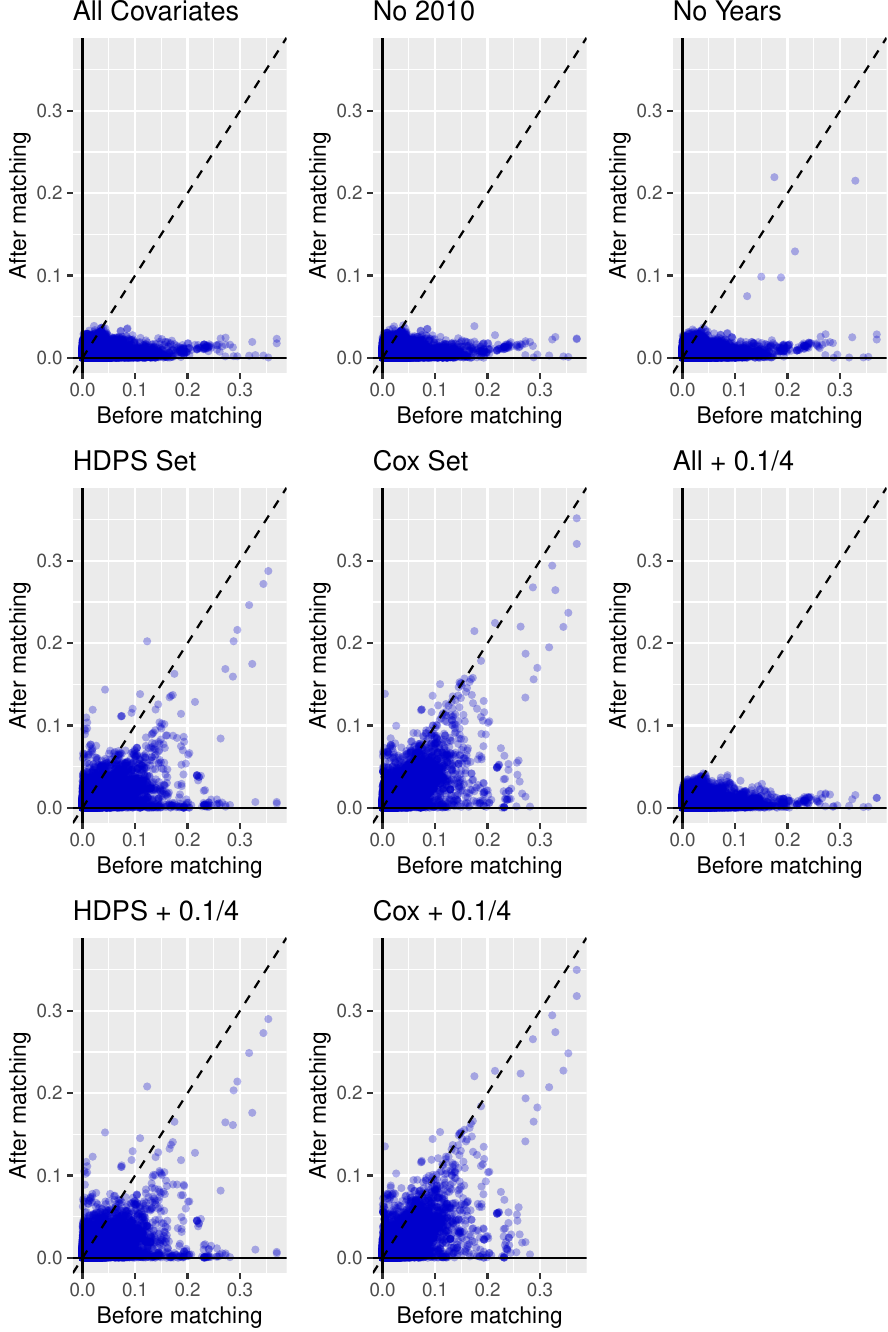}
  	\caption{Pre-matching vs. post-matching covariate absolute standardized differences for All Covariates. Each point represents one covariate. The three plots from simulated IV PS models are taken from simulations with 10\% IV prevalence and relative risk of 4.}
	\label{fig: balance_all}
\end{figure}

\begin{figure}[H]
	\centering
  	\captionsetup{width=\textwidth}
  	\includegraphics[width=\textwidth, height=1.25\textwidth] {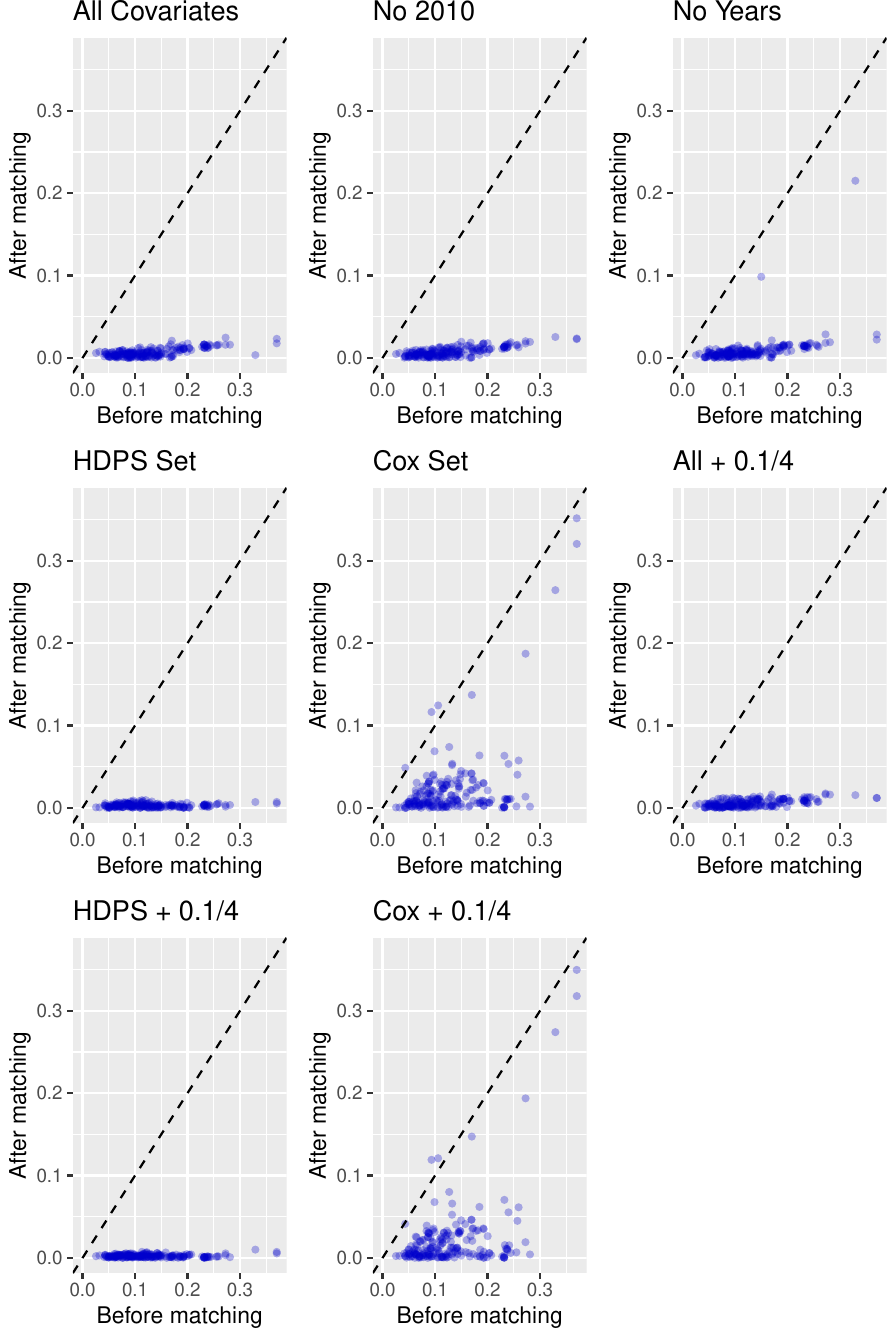}
  	\caption{Pre-matching vs. post-matching covariate absolute standardized differences for HDPS Set covariates. Each point represents one covariate. The three plots from simulated IV PS models are taken from simulations with 10\% IV prevalence and relative risk of 4.}
	\label{fig: balance_hdps}
\end{figure}

Propensity scores reduce confounding by creating comparable cohorts that are balanced with respect to pretreatment covariates.
Figure \ref{fig: balance_all} shows the covariate balance for the All Covariates set of covariates.
The All Covariates PS model does the best in balancing the covariates, and removing calendar year 2010 or all calendar years from the PS model results in the respective calendar years becoming unbalanced.
The HDPS Set PS model does a poor job with covariate balance, and the Cox Set PS model does an even worse job.
Adding a strong IV to the All Covariates, HDPS Set, and Cox Set PS models has very little effect on the covariate balance distribution, even though we have seen it has a strong effect on the PS distribution (Figure \ref{fig: ps_plots}).
Figure \ref{fig: balance_hdps} shows the covariate balance of just the HDPS Set covariates.
The HDPS Set PS model does the best at balancing covariates, and the All Covariates PS model also keeps all after-matching standardized differences below 0.05.
However, the Cox Set PS model fares poorly on these covariates' balance, and fails to balance numerous covariates.

\begin{figure}[H]
	\centering
  	\captionsetup{width=\textwidth}
  	\includegraphics[width=\textwidth, height=1.25\textwidth] {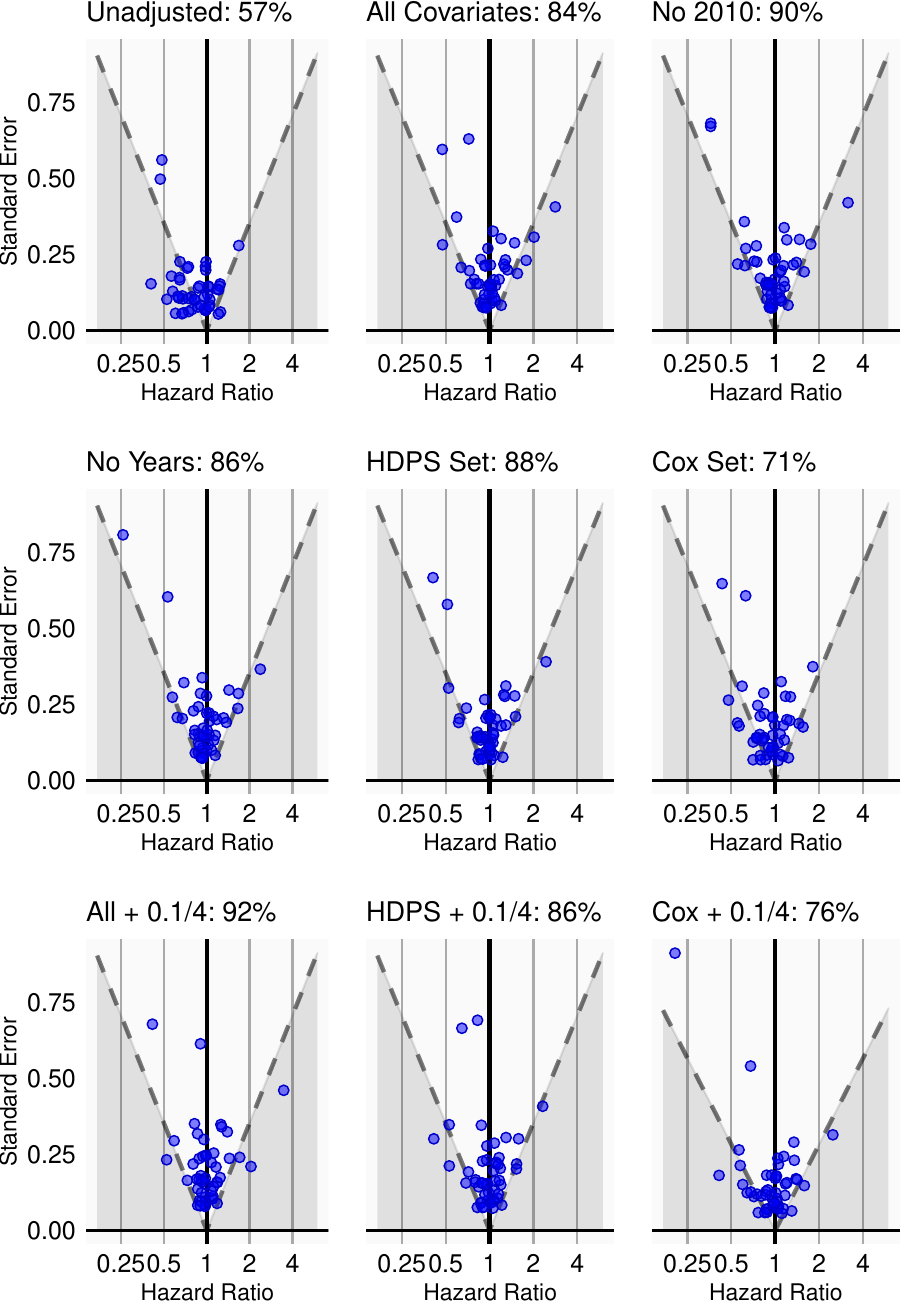}
  	\caption{Negative control outcome estimates with associated coverage of presumed true hazard ratio of 1. Each point represents one negative control estimate. Estimates above the dotted lines include 1 in their 95\% confidence intervals and are not statistically significant, while estimates below the dotted lines no not include 1 and are statistically significant. }
	\label{fig: neg_controls}
\end{figure}

Figure \ref{fig: neg_controls} shows the negative control outcome estimates generated using the PS models, along with the coverage by the estimates of the presumed true hazard ratio of 1.
Nominally, 95\% of the estimates' 95\% confidence intervals should include 1.
The unadjusted estimates have the smallest coverage, and have a mean estimate that is noticeably negative.
At 84\%, the All Covariates PS model has similar coverage compared to the HDPS Set at 88\%, while both are higher than he Cox Set at 71\%.
Removing calendar year 2010 and all calendar years both increase the coverage from the All Covariates PS model.
Adding a simulated IV to the All Covariates, HDPS Set, and Cox Set PS models increases the number of negative control estimates that produce nonsignificant confidence intervals.

\section{Discussion}

The propensity score is defined as an estimate of treatment assignment probability using pretreatment covariates, and its use is sufficient in removing bias from all measured confounders \cite{rosenbaum1983central,imbens2015causal}.
In other words, knowing the true treatment assignment probability would allow for perfectly unbiased outcome effect estimates in observational studies, through the design of stratified studies \cite{rubin2007design}.
Knowing this, it makes definitional sense to build the PS model with the goal of treatment prediction, and include only pretreatment covariates; the outcome, which postdate the treatment, would have no role in the PS model \cite{rubin2008objective,rubin2009should}.
Large scale regularized regressions are a natural approach to estimating the PS in the presence of thousands of covariates, as are available in longitudinal observational databases \cite{tian2018evaluating}.

Instrumental variables are established to be bias amplifiers in numerous theoretical and simulation studies \cite{ding2017instrumental,wooldridge2016should,bhattacharya2007instrumental,lefebvre2008impact,myers2011effects}.
There seems to be agreement that known IVs should be removed from the set of conditioning variables \cite{vanderweele2019principles}.
The question becomes, how can we identify IVs in real-world observational data that adhere to multiple unverifiable causal criteria?
Faced with this dilemma, some authors have argued that despite its definition, the purpose of the PS is as an adjustment to reduce estimation bias, and advocate for including covariates based on association with outcome \cite{patrick2011implications}.
This approach embraces a novel ideology for PS estimation: fit an outcome model, use the results to build a PS, and use the PS in another outcome model.
The HDPS utilizes a univariate screen to identify the most outcome-associated covariates for PS model inclusion \cite{schneeweiss2009high}, and has become a common tool for automated PS model construction \cite{rassen2011covariate,schneeweiss2018automated}.

We observe that sacrificing treatment prediction to create PS model covariate sets based on outcome association has expected consequences in PS distribution quality, as visualized through preference scores that normalize propensity scores by the treatment prevalence \cite{walker2013tool}.
Larger PS models do an increasingly better job at separating PS distributions of the target and comparator treatment (Figure \ref{fig: ps_plots}).
The All Covariates PS model, built on thousands of covariates, achieves the most separation between dabigatran and warfarin populations, while leaving enough overlap to allow for meaningful comparison between PS matched or stratified subjects.
In contrast, the HDPS Set PS model, built on 200 covariates identified through a univariate screen for outcome association, does not identify a large set of warfarin patients with low preference score.
The Cox Set PS model, built on 74 covariates identified in a multivariate regularized outcome model, has very little preference score separation between the two groups.
For all three PS models, inclusion of a simulated IV creates a spike in the preference score distribution at the high end of the scale near 1, showing that indeed the simulated IV strongly affects treatment assignment probability.

One might presume that a simulated IV with a strong effect on the PS distribution would diminish the relative contribution of other covariates in the PS model and have a detrimental effect on covariate balance.
Surprisingly, we observe that inclusion of a simulated IV has almost no perceptible effect on covariate balance (Figures \ref{fig: balance_all} and \ref{fig: balance_hdps}).
We do observe a similar relationship between PS model size and covariate balancing performance: the larger the PS model, the more covariates are successfully balanced (Figure \ref{fig: balance_all}).
The All Covariates PS model is built on all covariates, and even though the resultant PS model only has 900 nonzero coefficients, all covariates are satisfactorily balanced.
The HDPS Set and Cox Set PS models fail to restrict all after-matching standardized differences to below 0.1.
When we observe the covariate balance only for the 200 HDPS Set covariates, the HDPS Set PS model unsurprisingly performs the best covariate balancing (Figure \ref{fig: balance_hdps}).
However, the All Covariates PS model also performs excellently, showing that including [even vastly] more PS model covariates does not compromise the balance of a smaller subset of covariates that may be of interest to the investigator.

In our plasmode simulations under a known true hazard ratio, the Cox Set PS model unsurprisingly demonstrates the least bias, as it builds the PS using the exact covariates used to build the parametric outcome generating model (Figure \ref{fig: sim_bias4} and Table \ref{table: bias4}).
Despite doing a poorer job balancing the 74 covariates of the Cox Set that were used for outcome simulation, the HDPS Set PS model has slightly smaller bias than the All Covariates PS model.
The HDPS Set PS model also has fewer covariates, 26, that overlap with the Cox Set than the All Covariates PS model, at 31.
These results suggest that there is merit to selecting PS model covariates by outcome association when it comes to the ultimate goal of study bias.
Interestingly, removal of our suspected IV calendar year 2010 from the All Covariates PS model increases -- rather than decreases -- study bias, and removal of all calendar years further increases the bias.
Additionally, inclusion of a simulated IV to the All Covariates and Cox Set PS models often decreases the study bias.
Most strikingly, the All Covariates PS models with simulated IV have smaller bias with stronger and more prevalent simulated IV.
Inclusion of simulated IV does seem to generally increase variance across observed PS models.

Instrumental variables are widely known as bias amplifiers \cite{ding2017instrumental}, yet our simulation results show them having the opposite effect: removing suspected calendar year IVs slightly increases bias, while adding a simulated IV sometimes decreases bias.
We notice that published simulation studies utilize small simulation models in which the IV is one of a few -- if not the only -- simulated covariates \cite{bhattacharya2007instrumental,lefebvre2008impact,myers2011effects}.
Meanwhile, we are adding simulated IVs (or removing suspected IVs) from much larger models with at least 74 covariates and up to tens of thousands of covariates.
Our large PS models more accurately reflect real-world scenarios in which longitudinal observational databases provide many thousands of potential confounding covariates.
While we cannot explain the observed paradoxical IV effects, we believe that IVs have much weaker effect in real-world data than in small simulations.
A similar view, that adjusting for a suspected (and possibly imperfect) IV likely reduces net bias, is shared by one of the aforementioned simulation studies \cite{myers2011effects}.

Our plasmode simulations reveal somewhat of a circular result: using the exact covariates that affect the outcome in the PS model produces almost no study bias.
Unfortunately, it is impossible to know the exact outcome generating process of real-world outcomes of interest.
Whether through univariate screens or multivariate regressions, whatever outcome models we utilize to select covariates are inherently parametric and likely fail to capture the ``true" generative model.
For this reason, results from simulation experiments often fail to translate to real-world applications.
Additionally, real-world outcomes of interest are often rare, limiting our ability to fit accurate outcome models with multiple covariates.

In contrast, negative controls are able to provide what simulation experiments cannot -- a standard of clinical truth (that of no effect) in real-world data.
By using real data as negative control outcomes, our negative control experiments are able to estimate the distribution of residual study bias.
Our negative control experiments show a clear result: a PS model based on modeling treatment assignment results in less residual study bias as measured by negative control distributions (Figure \ref{fig: neg_bias} and Table \ref{table: neg_dist}).
The All Covariates PS model and associated simulated IV PS models have smaller residual bias, smaller variance, and higher coverage than the respective HDPS Set PS models, which in turn perform better than the Cox Set PS models.
Inclusion of simulated IV does seem to generally increase negative control distribution variance across observed PS models.

We offer a word of a caution in selecting covariates based on outcome association through the HDPS. 
The HDPS only screens the 500 covariates that have the highest prevalence, and does not screen low-prevalence covariates \cite{schneeweiss2009high}.
When we do include all covariates in the univariate screen, the resultant PS model fails to separate the treatment and comparator cohorts.
Furthermore, both the plasmode simulation bias and the residual study bias are substantially larger than that of other methods.
This is due to low prevalence covariates dominating the ranked list of covariates by apparent relative risk.
Our use of regularized regression \cite{tian2018evaluating} to fit PS models avoids this issue in the All Covariates PS model despite inclusion of all covariates, as low-prevalence covariates are shrunk by the $L_1$ penalty \cite{tibshirani1996regression} to have zero coefficients.

In conclusion, we find that simulated IVs have at most a weak effect on bias in simulations and negative control experiments based on large-scale real-world data.
We do find evidence in negative control experiments that removing suspected IVs increases precision but worsens bias.
IVs also have very little effect on covariate balance despite strongly affecting PS distributions.
If we were privy to real-world causal models, our simulation results suggest using outcome sensitive PS models for smallest estimation bias.
Unfortunately, real-world data cannot be omnisciently modeled, and the results of simulation experiments may not offer practical comparisons of PS methods.
Instead, we prefer to conduct negative control experiments that approximate residual bias, and those results confirm that large PS models curated through regularized regression \cite{tian2018evaluating} offer the least bias with or without instrumental variables.

\section*{Acknowledgments}

This work was partially supported through National Institutes of Health grants F31LM012636 and R01LM006910,
Food and Drug Administration's Biologics Effectiveness and Safety System contract,
and Australian National Health and Medical Research Council grant GNT1157506.
We thank NVIDIA Corporation for providing massive parallelization resources.

\bibliographystyle{myBibStyle}
\bibliography{iv_refs}

\end {document}